\documentclass[runningheads]{llncs}
\usepackage{graphicx}

\usepackage{multirow}
\usepackage{tabularx}
\usepackage{booktabs}

\usepackage[table]{xcolor}
\usepackage{hyperref}
 \usepackage{glossaries}
\begin{document}
\title{Transesophageal Echocardiography Generation using Anatomical Models}
%
%

\author{Emmanuel Oladokun\inst{1} \and
Musa Abdulkareem\inst{2} \and
Jurica Šprem\inst{2} \and
Vicente Grau\inst{1}}

\authorrunning{E. Oladokun et al.}
\institute{University of Oxford, Institute of Biomedical Engineering \and
GE HealthCare, Cardiovascular Ultrasound R\&D
\email{emmanuel.oladokun@eng.ox.ac.uk}}

\maketitle              
\begin{abstract}
Through automation, deep learning (DL) can enhance the analysis of transesophageal echocardiography (TEE) images. However, DL methods require large amounts of high-quality data to produce accurate results, which is difficult to satisfy. Data augmentation is commonly used to tackle this issue. In this work, we develop a  pipeline to generate synthetic TEE images and corresponding semantic labels.  The proposed data generation pipeline expands on an existing pipeline that generates synthetic transthoracic echocardiography images by transforming slices from anatomical models into synthetic images. We also demonstrate that such images can improve DL network performance through a left-ventricle semantic segmentation task. For the pipeline's unpaired image-to-image (I2I) translation section, we explore two generative methods: CycleGAN and contrastive unpaired translation. Next, we evaluate the synthetic images quantitatively using the Fréchet Inception Distance (FID) Score and qualitatively through a human perception quiz involving expert cardiologists and the average researcher. 

In this study, we achieve a dice score improvement of up to 10\% when we augment datasets with our synthetic images. Furthermore, we compare established methods of assessing unpaired I2I translation and observe a disagreement when evaluating the synthetic images. Finally, we see which metric better predicts the generated data's efficacy when used for data augmentation.

\end{abstract}

\section{Introduction}
Medical imaging is an essential tool in cardiovascular medicine. Its use can be found in screening, early diagnosis, treatment, and follow-up. Echocardiography (Echo) is a key contributor to the assessment and management of cardiac diseases \cite{Potter2019TheCare}. Transesophageal Echocardiography (TEE), specifically, is an invasive form of echo that provides additional and relatively clear visualisation of the heart \cite{Hahn2013ASEAnesthesiologists}. Due to its invasive nature, it is uncommonly used and researched relative to other forms, e.g. transthoracic echocardiography (TTE). TEE provides information that is key for diagnosing cardiac pathologies and assessing cardiac performance, including calculating quantitative metrics. 

The acquisition process for relevant physiological variables would benefit from automation, which could be accomplished using state-of-the-art computer vision methods. Deep learning (DL) methods have been proven effective in modern image analysis, especially Convolutional Neural Networks (CNNs) \cite{Chi2017ThyroidNetwork,PM2017TransferImages,Hafiane2017DeepImages,Li2019OverfittingSegmentation}. However, the large amount of high-quality labelled data generally required for training, is a challenge. Medical professionals are required to generate the labels (i.e., gold standard approximations to the ground truth), which is often expensive and time-consuming. Furthermore, it has been shown that there is significant inter-observer variability in the labelling of echo images by experienced cardiologists, resulting in up to 22\% variability in the calculation of physiological variables \cite{AC2015QualityExperience,A2010ReproducibilityStudy}. Differences occur in the labels because echo images are noisy and often contain artefacts. As a result, boundaries delineating structures are blurred and become open to interpretation. Finally, updating manual labels in light of new information is also time-consuming. In addition to these difficulties, acquiring large amounts of TEE data is especially difficult due to its relatively infrequent use.

Datasets of inadequate size and quality are a common problem in DL, motivating many attempts at finding a solution. A common approach is data augmentation, which aims to increase the size and diversity of a dataset. Most notably, Ronneberger et al. \cite{Ronneberger2015U-Net:Segmentation} showed that successfully training a deep CNN for semantic segmentation with a small, labelled set of images is possible by randomly applying augmentations, such as rotation and scaling, to existing data to generate 'new' data. However, traditional data augmentation methods are restricted by the strong correlation between the original and augmented samples.  
Rather than only using real datasets, some researchers have resorted to generating synthetic datasets using synthetic augmentation methods. These methods generally fall into two categories namely, same-domain and cross-domain synthesis. The former involves creating data in the same domain, and the latter involves using data from a source domain to create data in a target domain. The data synthesis approach has largely been spurred on by advancement in generative adversarial networks (GANs), especially the advent of the CycleGAN \cite{Zhu2017UnpairedNetworks}. For example, conditional GANs (cGANs) have been used to synthesise X-ray images \cite{Tang2019XLSor:Generation}, and pairs of MRI and CT images were synthesised using a CycleGAN \cite{ChartsiasAdversarialData}. Echo images are considerably more difficult to synthesise due to the complex speckle patterns and numerous artefacts. Attempts to synthesise echo images have traditionally taken a modelling approach where the physics behind the observed noise patterns are simulated \cite{Alessandrini2015ADatabase}. Unfortunately, this approach often lacks realism and has poor scalability due to the large computational power needed to run the simulations. Abdi et al. \cite{Abdi2019GAN-enhancedGeneration} successfully took a synthetic augmentation approach to synthesise echo images by training a patch-based cGAN, but did so using a paired, scarcely annotated dataset.

 Using detailed 3D anatomical models and a CycleGAN, Gilbert et al. \cite{Gilbert2021GeneratingSegmentation} developed a pipeline where synthetic images are generated from existing high-quality annotations. Using this approach, they generated a synthetic labelled dataset of echo images and trained a CNN for left ventricle (LV) and left atrium segmentation. However, their example only presents two views, apical 4-chamber and apical 2-chamber, which are characteristic views of transthoracic echo. The TEE modality, however, is more data-deprived due to its invasive nature and the large variety of acquired views,  making dataset acquisition more challenging. Since the inception of \cite{Gilbert2021GeneratingSegmentation}, newer methods for unpaired image-to-image (I2I) translation have been developed, such as contrastive unpaired translation (CUT) \cite{Park2020ContrastiveTranslation}. Unlike the CycleGAN, CUT is one-sided and does not require as many auxiliary networks and loss functions. Consequently, it has lower memory requirements and requires fewer computational resources to train. Now, with other promising unpaired I2I methods available, we explore whether CUT improves upon the CycleGAN in this context. Furthermore, it is desirable to have a metric that predicts whether the synthetic images from a generator will improve segmentation performance without having to train a segmentation network on each augmented dataset to compare. This contrasts with current metrics for predicting augmentation impact that require training networks to perform the downstream task (segmentation in this case) using clean and augmented datasets \cite{Gontijo-Lopes2020AffinityAugmentation}, which is unrealistic for GAN training. Therefore, we also investigate whether the metrics used to evaluate I2I translation methods can predict augmentation impact. The contributions of this paper are as follows:
\begin{enumerate}
\item	For the first time, we develop a pipeline\footnote{Code is available at \url{https://github.com/adgilbert/pseudo-image-extraction.git}} capable of automatically generating realistic, labelled synthetic images showing 19\footnote{Details of which 19 views can be found in the supplementary material.} standard views of TEE (as defined by the American Society of Echocardiography \cite{Hahn2013ASEAnesthesiologists}). 
\item	We show that these synthetic images can be used to improve performance when tested on a LV segmentation task
\item  For the first time, we investigate the link between Fréchet Inception Distance (FID) Score \cite{HeuselGANsEquilibrium} and human-judged realism in synthetic TEE images and explore if either metric can also predict augmentation benefit
\end{enumerate}
\section{Methods}
\begin{figure}
\includegraphics[width=\textwidth]{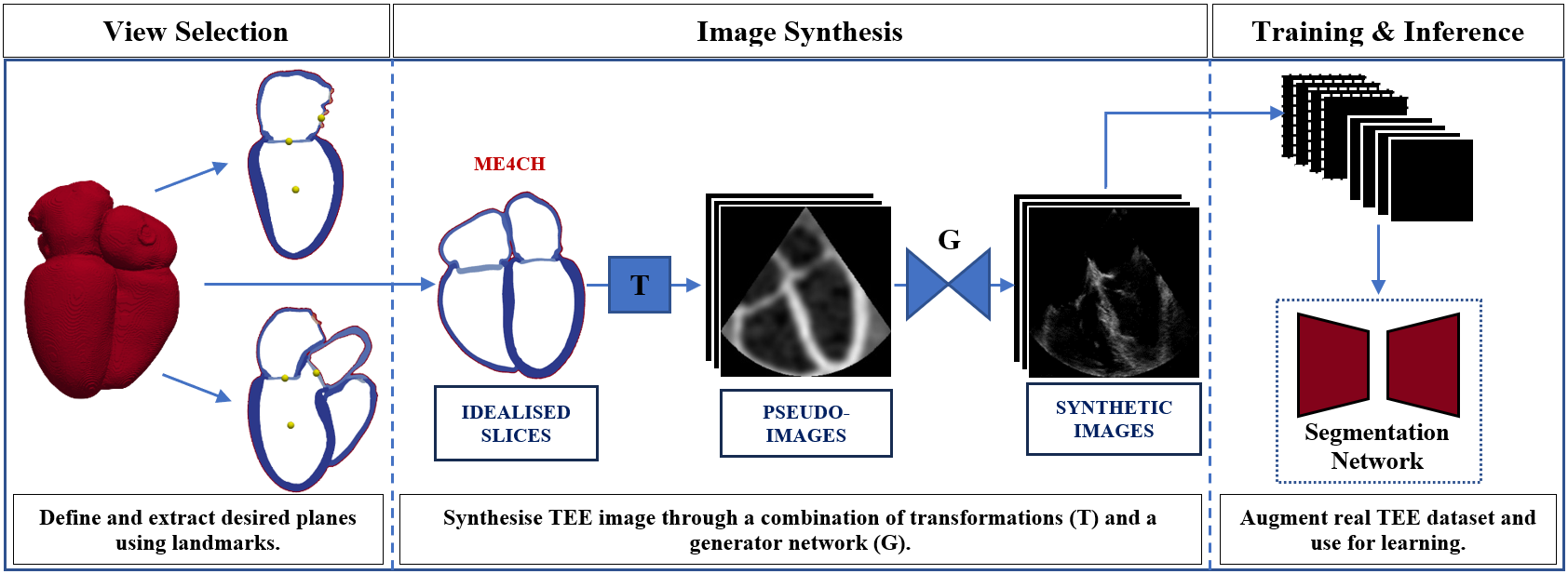}
\caption{\textbf{Synthetic TEE Generation and Image Segmentation Pipeline} Using landmarks, desired TEE planes are extracted from each heart model. These ideal slices are the ground truth labels for the synthetic images. Next, pseudo-images are made by adding the acquisition cone and some transformations e.g. Gaussian blurring, shadowing etc., The image synthesis phase concludes with pseudo-images being passed through a trained generator. The synthetic images and their masks can then be used to augment a real dataset for the chosen task}
\label{fig:pipeline_fig}
\end{figure}

\subsection{Pseudo-image Generation}

\subsubsection{Anatomical Models} 
Figure \ref{fig:pipeline_fig} shows the journey from anatomical models to usable synthetic images and eventually, augmentation. These models were generated using CT images of asymptomatic hearts acquired at end-diastole. Firstly, the main heart structures were automatically segmented and post-processed, then a mesh was constructed following methods described in \cite{Rodero2021LinkingHeart}. The models contain a full set of tissue labels and labels for all four chambers. The major vessels: aorta, pulmonary veins, pulmonary arteries, and both venae cavae are also labelled. More information on the model creation process can be found in the following works \cite{Rodero2021LinkingHeart,Strocchi2020ASimulations}. From a set of 19 subject-specific anatomical models, Principal Component Analysis was used to create a statistical model. We sampled from this to generate 99 models, which we then used to generate pseudo-images. Full details of this expansion can be found in \cite{Gilbert2021GeneratingSegmentation}.
\subsubsection{TEE View Extraction} 
According to the American Society of Echocardiography \cite{Hahn2013ASEAnesthesiologists}, there are 28 standard views acquired in a comprehensive TEE examination. These views can be defined by prescribing which structures should be visible in the echo images. 
Figure \ref{fig:pipeline_fig} shows how we create pseudo-images for specific views (e.g. Mid-Esophageal 4-Chamber (ME4CH)) extending on the method described in \cite{Gilbert2021GeneratingSegmentation}. Three structures that should be present in the image are selected and used to define a unique slice plane. In the case of ME4CH, the three landmarks are the centres of mass of the mitral valve, tricuspid valve, and LV. Furthermore, to mimic the natural variations in ultrasounds during acquisition, we rotate the extracted slice about two mutually perpendicular axes to produce images where only some landmarks are in-plane. The rotation axes were defined by using two landmarks to generate a line along the long- or short-axis of the heart. 
This process was repeated with different heart models, and the results were visually assessed to ensure the extracted slice was similar to the expected view, despite variations in the heart morphology. Following the slice extraction, an ultrasound cone mask, random noise, shadows, and blurring are added to the image to complete the pseudo-image generation. The parameters of these transformations are randomly sampled from a predefined range to introduce more variability (details on exact parameters can be seen in the GitHub repository) 
Using this approach, we extended the pipeline by providing functionality to generate 19 of the possible 28 standard views. Our current methodology cannot generate the remaining 9 views, either because the structures present in a view were not available in the shape models or because there were insufficient landmarks to define the plane robustly. 

\subsection{Image Synthesis}
\subsubsection{Unpaired I2I Models}
\label{sec:i2i}
Given unpaired datasets $X$ (source domain), and $Y$ (target domain), both the CUT \cite{Park2020ContrastiveTranslation} and CycleGAN \cite{Zhu2017UnpairedNetworks} methods attempt to find an optimal generator $\hat{G}$ that best maps from $ X \rightarrow Y$. The CycleGAN's two generators $\{G, F\}$ try and fool their respective discriminators $\{D_{X}, D_{Y}\}$.  High-quality mappings can be attained by combining these networks and adding a cyclic-consistency loss. The CycleGAN’s loss function is 
\begin{equation}
\begin{aligned}
        \label{eq:cyclegan_obj}
     \mathcal{L}(G,F,D_{X},D_{Y}) = \, &\mathcal{L}_{adv}(G,D_{Y},X,Y) \,+ \mathcal{L}_{adv}(F,D_{X},Y,X) \,+ \\ 
     &\lambda_{cyc}\mathcal{L}_{cyc}(G,F) \,+ \lambda_{idt}\mathcal{L}_{idt}(G,F) ,
\end{aligned}
\end{equation}
where $\mathcal{L_{\{\text{adv, cyc, idt}\}}}$ corresponds to the adversarial, cyclic and identity losses, and $\lambda_{\{cyc,idt\}}$ allow the weighting of individual loss terms.

CUT takes a different approach and leverages contrastive learning to encourage content preservation while allowing changes in appearance.  CUT samples patches from the original and generated images then via a projection head $H$, maps patches in the same position together in latent spaces and distances patches from different positions. Its loss function is
\begin{equation}
\begin{aligned}
    \label{eq:cut_loss_func}
    \mathcal{L}(G,H,D) = \, &\mathcal{L}_{adv}(G,D,X,Y) + \lambda_{x}\mathcal{L}_{PatchNCE}(G,H,X) + \\
    & \lambda_{y}\mathcal{L}_{PatchNCE}(G,H,Y),
\end{aligned}
\end{equation}
where  $\mathcal{L}_{PatchNCE}$ is the contrastive loss used for the patches, and $\lambda_{\{x,y\}}$ are weighting hyperparameters

\subsubsection{Data}
The set of real ME4CH echos used in this study consists of 2,914 images sourced from 26 subjects. The images are sampled freely from all parts of the cardiac cycle and were sourced from several institutions in different countries by GE HealthCare. Splits were made on a subject basis. Images from 12 subjects were removed to generate $R_{I2I}$, which contains 1,959 images. From the remaining 14 subjects, 182 images were sampled, labelled, and then split into train ($R_{train}$) and test ($R_{test}$) sets of sizes 155 and 27, respectively. The labelling was performed by two expert cardiologists who use echo daily.

Using the extended pipeline, we generated 854 pseudo-images and the corresponding labels of the ME4CH plane. This dataset was split into sets $P_{I2I}$ and $P_{seg}$, each containing 503 and 351 images, respectively. 

\subsubsection{Generative Training}
To train the generators discussed in "Unpaired I2I Models", we made use of $P_{I2I}$ and $R_{I2I}$ as the source domain and target domain data. At train time, we manually review and sample results every five epochs. The best-performing generators in terms of realism were selected by the first author, giving generators $\hat{G}_{cyc}$ and $\hat{G}_{cut}$ (Details of networks, hyperparameters and training schedules can be seen in the supplementary Table 2). Next, synthetic datasets were generated by passing $P_{seg}$ through each generator, thereby generating synthetic sets $\hat{S}_{cyc}$ and  $\hat{S}_{cut}$ which both contain 351 images.

\subsubsection{Evaluation}
The synthetic images were then evaluated in three separate ways. 
To test realism, we conducted a Turing-like test where two expert cardiologists and six researchers in our group were tasked with labelling echos as real or synthetic. Using VGG's Image Annotator \cite{Dutta2019TheVideo}, we developed a User Interface (UI) to show the participant an image and ask whether they believed it was real or synthetic. Before the test began, participants were shown several real ME4CH echos. The user could then toggle back and forth through the images, providing responses. The quiz dataset comprised 120 images, 60 real, 30 from CUT, and 30 from CycleGAN.
Secondly, we calculate each set's FID score, which quantifies the similarity between two datasets and is commonly used in I2I translation literature \cite{MaximilianSeitzer2020Pytorch-fid:PyTorch}.

Finally, to evaluate the synthetic data augmentation impact, we use the trusted nnunet framework \cite{Isensee2018NnU-Net:Segmentation} to train U-Nets for segmenting the LV.  Firstly, we generate baselines by training nnunet models on datasets containing a randomly sampled percentage of the original train set. Then, we independently add $\hat{S}_{cyc}$ and  $\hat{S}_{cut}$ to the real image sets, giving: purely synthetic datasets, mixed datasets with varying fractions of the real images, and a set containing all real images and all synthetic images from a particular generator. We term the performance metric difference between the real baseline and the augmented sets the 'delta $(\Delta)$ metric'. In all cases, we used 3-fold cross-validation and ensured only real images were present in the validation set, thereby only evaluating performance on real images. Standard shape and texture augmentations were included for all runs as they are cheap to run, and we believe the addition of synthetic images can improve performance alongside such transformations.
\section{Results \& Discussion}
\subsubsection{Generative Results}
\begin{table}[h]
\centering
\caption{Participant results on the quiz with expert performance and researcher performance separated. Each person was shown 60 real (R) and 60 synthetic (S) images. { \it *Accuracy was rounded to 1 d.p whilst frequencies were rounded to the nearest integer.  ** 95\% confidence interval}}
\label{tab:quiz_results}
\begin{tabularx}{\textwidth}{
>{\hsize=1.3\hsize\textwidth=\hsize}X 
>{\hsize=1.3\hsize\textwidth=\hsize}X
>{\hsize=0.88\hsize\textwidth=\hsize}X
>{\hsize=0.88\hsize\textwidth=\hsize}X
>{\hsize=0.88\hsize\textwidth=\hsize}X
>{\hsize=0.88\hsize\textwidth=\hsize}X
>{\hsize=0.88\hsize\textwidth=\hsize}X
>{\hsize=0.88\hsize\textwidth=\hsize}X}

\hline
\multicolumn{7}{c}{\textbf{Human Perception Quiz}} \\ \hline
\multirow{2}{*}{Participant} &
  \multirow{2}{*}{Accuracy {[}\%{]}} &
  \multirow{2}{*}{R as R} &
  \multirow{2}{*}{R as S} &
  \multirow{2}{*}{S as R} &
  \multirow{2}{*}{S as S} &
  \multirow{2}{*}{F1-Score} \\
            &         &       &      &      &     &       \\ \hline
Expert 1    & 95.0    & 55    & 5    & 1    & 59  & 94.8 \\
Expert 2    & 79.2    & 60    & 0    & 25   & 35  & 82.8 \\ \hline
\multirow{2}{*}{Researchers*} &
  \multirow{2}{*}{\begin{tabular}[c]{@{}l@{}}69.7 \\ $[60.2, 79.2]^{**}$\end{tabular}} &
  \multirow{2}{*}{39} &
  \multirow{2}{*}{21} &
  \multirow{2}{*}{15} &
  \multirow{2}{*}{45} &
  \multirow{2}{*}{68.4} \\
            &         &       &      &      &   &   \\ \hline
\end{tabularx}
\end{table}

Table \ref{tab:quiz_results} shows the confusion matrices for each expert and the researcher cohort. The experts were especially adept at identifying real images compared to the researchers. Moreover, the experts had significantly fewer false negatives (Real as Synthetic), showing strong knowledge of what TEE images should look like. Observing the F1-Score, a score which helps measure precision and recall simultaneously,  we see that the experts were noticeably better classifiers than the average researcher. This is despite Expert-2's accuracy being close to the average researcher's accuracy range.  Interestingly, Expert-1 performed extremely well in identifying the synthetic images as well. Consequently, we further investigated how they made their decisions. They commented that the real image's grey areas and black cavities were more homogeneous and that there were differences in the myocardium.
%
%
%
%

\begin{table}[h!]
\parbox{.44\textwidth}{
\caption{FID Score between an unseen set of TEE images and the synthetic sets. A lower score represents better fidelity between the generated and real images}
\label{tab:fid_score}

\centering
\begin{tabular}{cc}
\hline
\multicolumn{2}{c}{\textbf{FID Score $\downarrow$}}                                      \\ \hline
\multicolumn{1}{l}{\multirow{2}{*}{$\hat{G}_{cut}$}} & \multirow{2}{*}{$\hat{G}_{cyc}$} \\
\multicolumn{1}{l}{}                                 &                                  \\ \hline
\multicolumn{1}{l}{188}                              & 230                              \\ \hline
\end{tabular}

}
\parbox{.5\textwidth}{
\centering
\caption{Accuracy on the synthetic images only. A lower score shows more difficulty in identifying the generator's image as synthetic}
\label{tab:accuracy_by_gen}

\begin{tabular}{cccc}
\hline
\multicolumn{4}{c}{\textbf{Accuracy (\%) $\downarrow$}} \\
\hline 
 \multicolumn{2}{c}{$\hat{G}_{cut}$} & \multicolumn{2}{c}{$\hat{G}_{cyc}$} \\
\cmidrule(l){1-2} \cmidrule(l){3-4}
 Experts & Researchers & Experts & Researchers \\
\midrule
96.7 & 91.1 & 60.0 & 58.3 \\
\hline
\end{tabular}
}
\end{table}

Table \ref{tab:fid_score} and \ref{tab:accuracy_by_gen} show the FID score for each generator and how successful they were at fooling the quiz participants, respectively. Interestingly, the two metrics do not agree, which suggests that the FID score may not correlate well with human-judged realism, perhaps for medical images in general, but at least for ME4CH TEE images. In Table \ref{tab:accuracy_by_gen}, we see that $\hat{G}_{cyc}$ creates more realistic images compared to $\hat{G}_{cut}$. Furthermore, the expert and researcher cohorts performed similarly in identifying synthetic images from each generator, showing that expert opinion is not required to assess the quality of a generator's synthetic images properly.


\subsubsection{LV Segmentation}
In Table \ref{tab:lv_results}, we see the dice score achieved on the test set ($R_{test}$) by the models trained on datasets with differing mixtures of real and synthetic data. In all columns except $R_{train}^{20}$, the networks trained on mixed sets performed best, showing improvements of up to 10\%. The maximum dice score achieved (72.9) was by a network trained with mixed data and shows a 5.9\% improvement on the baseline. This is especially significant when considering the dataset's realistic size and its heterogeneity in time and source institutions. Moreover, this improvement occurred due to leveraging unlabelled data via the pipeline, and despite the synthetic images only depicting the heart at end-diastole, just like the anatomical models they were sourced from.

Interestingly, the models trained on $[\emptyset \cup R_{train}^{20}]$ and $[\hat{S}_{cut} \cup R_{train}^{20}]$ performed better than expected compared to some sets with more real data. One explanation for this could be the randomness of the data added from 20\% to 40\%. Adding training data only improves performance when the added data is relevant to the task at hand. Given the heterogeneity of the data, it is likely that the domain gap between datasets in the $R_{train}^{40}$ column and the test set is larger than the domain gap between the $R_{train}^{20}$ column and test set. This would likely lead to a drop in performance, as seen here. 
\begin{table}[]
\caption{ Dice score achieved on the test set. The column headers show real datasets with varying percentages of $R_{train}$, whilst the first column shows the source of the synthetic data and delta scores. Each non-grey element corresponds to the performance of an nnunet trained on the union of the column and row datasets ($<column> \bigcup <row>$). \textcolor{blue}{\textbf{Blue}} = Best Column Delta Score; \textbf{Bold =} Best Overall Dice Score}
\label{tab:lv_results}
\centering

\begin{tabularx}{\textwidth}{XXXXXXX}
\hline
\multicolumn{7}{c}{\textbf{Dice Score (x100) $\uparrow$}} \\ \hline
\multirow{2}{*}{} &
  \multirow{2}{*}{$\boldsymbol{\emptyset}$} &
  \multirow{2}{*}{$\boldsymbol{R_{train}^{20}}$} &
  \multirow{2}{*}{$\boldsymbol{R_{train}^{40}}$} &
  \multirow{2}{*}{$\boldsymbol{R_{train}^{60}}$} &
  \multirow{2}{*}{$\boldsymbol{R_{train}^{80}}$} &
  \multirow{2}{*}{$\boldsymbol{R_{train}}$} \\
                 &       &                                   &      &      &     &      \\ \hline
$\boldsymbol{\emptyset}$           & -     & 54.7                              & 47.9    & 51.1    & 53.5  &67.0   \\ 
$\boldsymbol{\hat{S}_{cut}}$  & 34.9  & 50.2                              & 48.2    & 49.5    & 63.6  & 69.5\\ 
\rowcolor[HTML]{dadada}
$\boldsymbol{\Delta_{cut}}$   & -     & \textcolor{blue}{\textbf{-4.5}}   & +0.3    & -1.6   & \textcolor{blue}{\textbf{+10.1}}  & +2.5   \\ 
$\boldsymbol{\hat{S}_{cyc}}$  & 20.9  & 44.2    & 50.4   & 53.7    & 61.8  & \textbf{72.9}   \\ 
\rowcolor[HTML]{dadada}
$\boldsymbol{\Delta_{cyc}}$   & -     & -10.5     & \textcolor{blue}{\textbf{+2.5}}    & \textcolor{blue}{\textbf{+2.6}}   & +8.3  & \textcolor{blue}{\textbf{+5.9}}   \\ \hline
\end{tabularx}
\end{table}

When we observe the generative model metrics in tandem with the segmentation results, we see that human realism is a better predictor for augmentation impact than FID Score. $\hat{S}_{cyc}$ produced the better delta score in more mixes and was responsible for the largest overall dice score. This agrees more with the quiz results; notably, the average researcher should be able to make this assessment. This disagreement in evaluation metrics is not unfeasible as even though FID Score usually correlates well with human perception, there is doubt as to whether this is true for medical images \cite{Bargsten2020SpeckleGAN:Processing}. The Inception Net used to calculate the FID Score is pre-trained on natural scene images. Unlike natural scene images, echo images can often hold valuable information in the noise and speckle patterns. Therefore, the FID Score may not be as meaningful for TEE images.

\section{Conclusion}
In this study, we extended the functionality of an existing pipeline to extract standard TEE views and generate their pseudo-images. We compared image translation methods for transforming pseudo-images into synthetic TEE images, evaluated them with the FID Score, and through a human perception quiz. Next, we showed that these synthetic images improved performance on an LV segmentation task by as much as 10\%. Finally, we observed that human perception and the FID Score disagreed in evaluating the realism of synthetic TEE images and that human perception was a better predictor of augmentation impact than the FID Score. Future work could entail exploring the use of these synthetic images for other tasks such as augmenting ultrasound datasets that are not TEE. Another useful path is the development of a quantitative I2I translation metric that can be trusted with medical images.
\subsubsection{Acknowledgements}
The authors thank D. Kulikova and A. Novikova for their help annotating images and participating in the quiz. We also thank the researchers who participated in the quiz.

%
%
%
\bibliographystyle{splncs04}
\bibliography{miccai2023}

\end{document}